\documentstyle[times,graphicx,amssymb,macros2e]{article}

\skip\footins 39pt plus 4pt minus 2pt
\def\footnoterule{\kern-19pt\hrule width.5in\kern18.6pt}%
\newcommand{\R}{{\mathbb{R}}}
\newcommand{\rmd}{{\mathrm{d}}}
\newcommand{\rme}{{\mathrm{e}}}
\newcommand{\half}{\frac{1}{2}}
\newcommand{\nn}{\nonumber}
\newcommand{\fig}[2]{\includegraphics[width=#1]{./figures/#2}}
\newcommand{\pfig}[2]{\parbox{#1}{\includegraphics[width=#1]{./figures/#2}}}

\newcommand{\bilderscale}{0.2}

\newlength{\bilderlength} 
\newcommand{\usebilderscale}{\bilderscale}
\newcommand{\bilderskip}{\hspace*{0.8ex}}

\newcommand{\diagram}[1]{%
\settowidth{\bilderlength}{\bilderskip%
\includegraphics[scale=\usebilderscale]{./figures/#1}\bilderskip}%
\parbox{\bilderlength}{\bilderskip%
\includegraphics[scale=\usebilderscale]{./figures/#1}\bilderskip}}

\begin{document}
\title{\bfseries \sffamily Why one needs a functional RG to survive in a 
disordered world}
\author{Kay J\"org Wiese\\
\small
Laboratoire de Physique Th\'eorique, Ecole Normale
Sup\'erieure, 24 rue Lhomond, 75005 Paris, France}

\maketitle

\abstract{In these proceedings, we discuss why Functional
Renormalization is an essential tool to treat strongly disordered
systems. More specifically we treat elastic manifolds in a disordered
environment.  These are goverened by a disorder distribution, which after
a finite renomalization becomes non-analytic, thus overcoming the
predictions of the seemingly exact dimensional reduction.  We discuss,
how a renormalizable field theory can be constructed, even beyond
2-loop order. We then consider an elastic manifold imbedded in $N$
dimensions, and give the exact solution for $N\to\infty$. This is
compared to predictions of the Gaussian replica variational ansatz,
using replica symmetry breaking. Finally, the effective action at
order $1/N$ is reported.}


\section{Introduction}\label{intro}

In these proceedings we consider an elastic manifold in a random
potential, as prototype for strongly disordered systems. Since for all
these systems temperature is irrelevant, we will only treat zero
temperature. The kind of systems we have in mind are domain walls in
dirty magnets, contact lines, charge density waves, vortex lattices,
to just mention a few. These results were obtained in collaboration
with Pierre Le Doussal
\cite{ChauveLeDoussalWiese2000a,LeDoussalWiese2001,LeDoussalWieseChauve2002,LeDoussalWieseChauve2003,LeDoussalWiese2002a,LeDoussalWiesePREPb,LeDoussalWiesePREPf,LeDoussalWiese2003a,RossoKrauthLeDoussalVannimenusWiese2003,LeDoussalWiese2003b,LeDoussalWiese2004a}. For
lack of space we restrict our discussion to the equilibrium.
Complementary material, especially for the depinning, can be found in
the earlier review \cite{Wiese2003a}.

\section{Physical realizations, model and observables}\label{model}
\begin{figure}[t]
\centerline{\fig{0.25\textwidth}{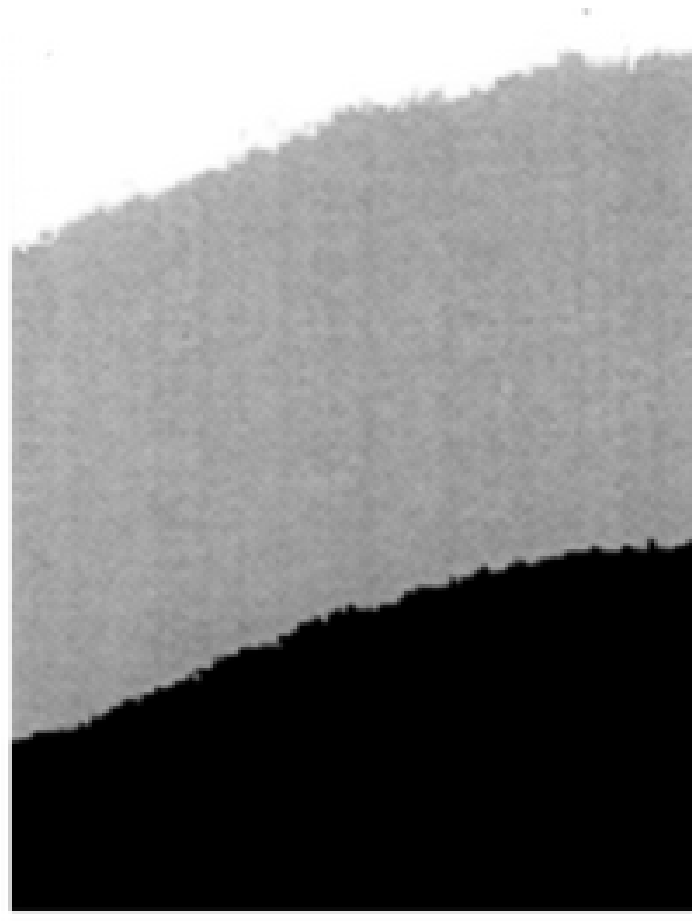}~~~\fig{0.7\textwidth}{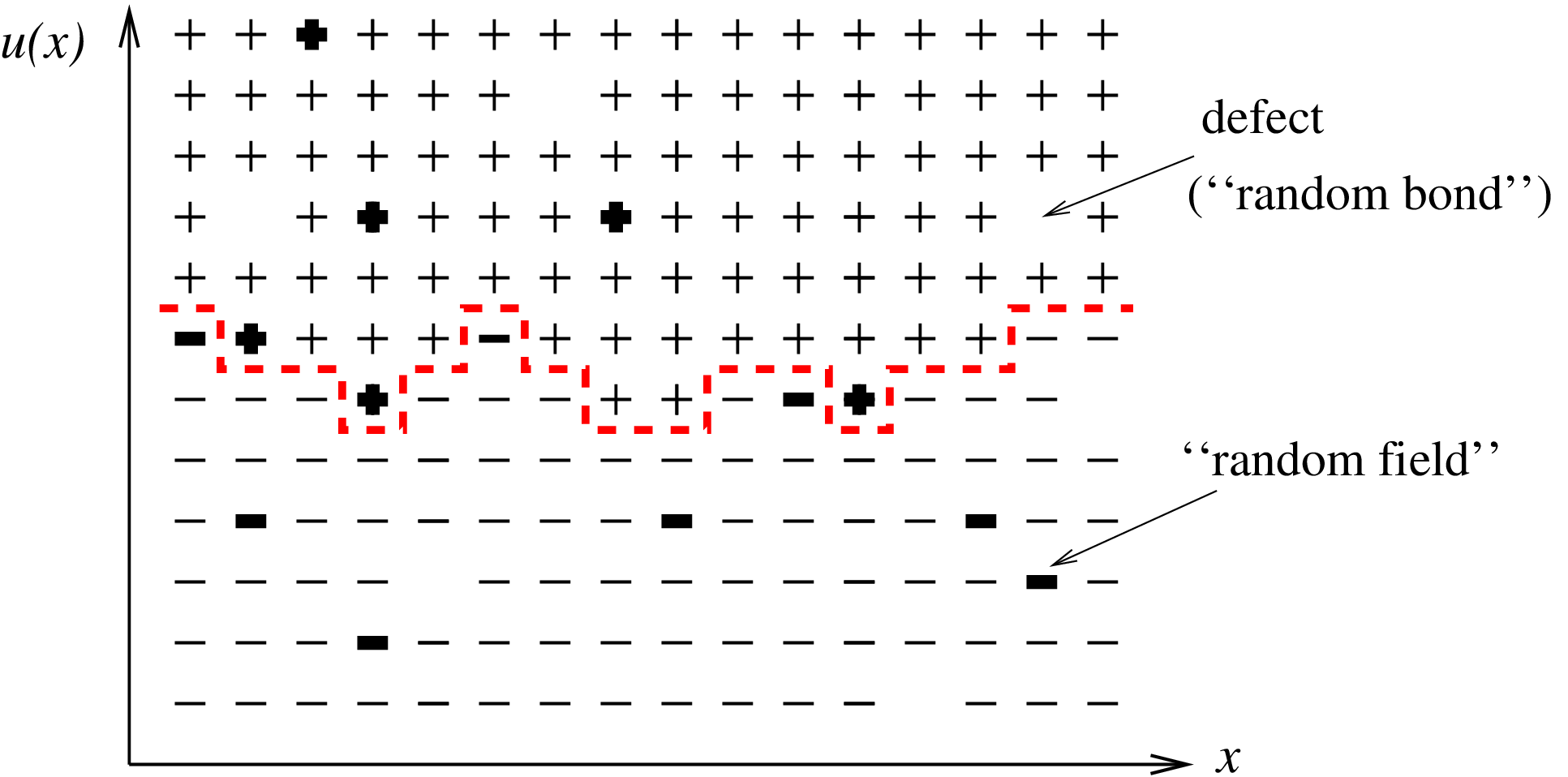}}
\caption{An Ising magnet at low temperatures forms a domain wall
described by a function $u (x)$ (right). An experiment on a thin
Cobalt film (left)
\protect\cite{LemerleFerreChappertMathetGiamarchiLeDoussal1998}; with
kind permission of the authors.}
\label{exp:Magnet}
\end{figure}
The simplest experimental realization is an Ising magnet. Imposing
boundary conditions with all spins up at the upper and all spins down
at the lower boundary (see figure 1), at low temperatures, a domain
wall separates a region with spin up from a region with spin down. In
a pure system at temperature $T=0$, this domain wall is completely
flat.  Disorder can deform the domain wall, making it eventually rough
again.  Figure 1 shows, how the domain wall is described by a
displacement field $u (x)$.  Another example is the contact line of
water (or liquid Helium), wetting a rough substrate. A realization
with a 2-parameter displacement field $\vec{u} (\vec x) $ is the
deformation of a vortex lattice: the position of each vortex is
deformed from $\vec x$ to $\vec x+ \vec u (\vec x)$.  A 3-dimensional
example are charge density waves.

All these models have in common, that they are described by a
displacement field $ x\in {\R}^d \ \longrightarrow\ {\vec u} (x) \in
\R^N$.  For simplicity, we set $N=1$, if not explicitly stated
otherwise.  After some initial coarse-graining, the energy ${\cal
H}={\cal H}_{\mathrm{el}}+{\cal H}_{\mathrm{DO}}$ consists out of two
parts: the elastic energy ${\cal H}_{\mathrm{el}}$ and the disorder
energy ${\cal H}_{\mathrm{DO}}$
\begin{equation}
{\cal H}_{\mathrm{el}}[u] = \int \rmd ^d x \, \half \left(
\nabla u (x)\right)^2 \ , \qquad {\cal H}_{\mathrm{DO}}[u] = \int \rmd
^{d} x \, V (x,u (x))\
\end{equation}
We choose the disorder at the microscopic scale Gaussian, with  correlations 
\begin{equation}\label{DOcorrelR}
\overline{V (u,x)V (u',x')} := \delta ^{d } (x-x') R (u-u')
\ .
\end{equation}
The most interesting observable is the roughness-exponent $\zeta $,
from the behavior of the correlation function
\begin{equation}\label{zetacor}
\overline{\left[ u (x)-u (y)\right]^{2}} \sim |x-y|^{2\zeta} \ .
\end{equation} 
Other observables are higher correlation functions or the free energy.

\section{Dimensional reduction} There is a beautiful and rather
mind-boggling theorem relating disordered systems to pure systems
(i.e.\ without disorder), which applies to a large class of systems,
e.g.\ random field systems and elastic manifolds in disorder. It is
called dimensional reduction and reads as
follows\cite{EfetovLarkin1977}:

\noindent {\underline{Theorem:}} {\em A $d$-dimensional disordered
system at zero temperature is equivalent to all orders in perturbation
theory to a pure system in $d-2$ dimensions at finite temperature. }

Let me give an example: The thermal expectation value for the 2-point
function scales as $\langle \left[ u (x) -u (y)\right]^{2} \rangle \sim
|x|^{2-d}$.  Making the dimensional shift implied by dimensional
reduction implies that the disorder-averaged 2-point function at zero
temperature is 
\begin{equation}\label{zetaDR}
\overline{\left[ u (x)-u (0) \right]^{2}} \sim x^{4-d} \equiv x^{2\zeta }
\quad \mbox{i.e.}\quad \zeta =\frac{4-d}{2}\ .
\end{equation}
We will see later that this is not true; but remains an important
benchmark due to fact that the ``theorem'' is correct to  all orders
in the disorder strength and its moments (i.e.\ when expanding in $R'' (0)$,
$R'''' (0)$, a.s.o.).  

\section{The Larkin-length}\label{Larkin} To understand the failure of
dimensional reduction, let us turn to an interesting argument given by
Larkin \cite{Larkin1970}. He considers a piece of an elastic manifold
of size $L$. If the disorder has correlation length $r$, and
characteristic potential energy $\bar \epsilon $, this piece will
typically see a potential energy of strength $E_{\mathrm{DO}} = \bar
\epsilon (\frac{L}{r} )^{{d}/{2}} $.  On the other
hand, there is an elastic energy, which scales like $E_{\mathrm{el}} =
c\, L^{d-2}$.  These energies are balanced at the {\em Larkin-length}
$L=L_{c}$ with $L_{c} = (\frac{c^{2}}{\bar \epsilon^{2}}r^{d})^{{1}/
( {4-d})}$.  More important than this value is the observation that in
all physically interesting dimensions $d<4$, and at scales $L>L_{c}$,
the membrane is pinned by disorder; whereas on small scales elastic
energy dominates. This means that $d=4$ is the upper critical
dimension.

\begin{figure}[b]
\centerline{\fig{13.4cm}{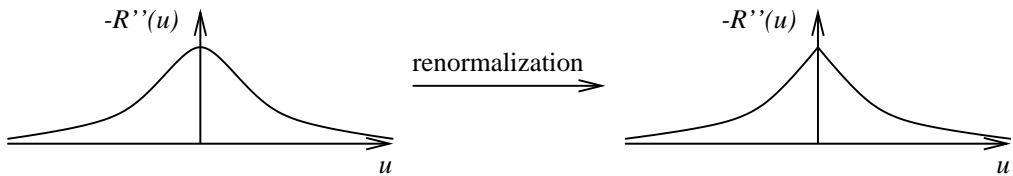}}
\caption{Change of $-R'' (u)$ under renormalization and formation of
the cusp.} \label{fig:cusp}
\end{figure}
\section{The functional renormalization group (FRG)}\label{FRG} Let us
now discuss a way out of the dilemma, posed by dimensional reduction:
We would like to make an $\epsilon =4-d$ expansion. On the other hand,
dimensional reduction tells us that the roughness is $\zeta
=\frac{4-d}{2}$ (see (\ref{zetaDR})). Even though this is
systematically wrong below four dimensions, it tells us correctly that
at the critical dimension $d=4$, where disorder is marginally
relevant, the field $u$ is dimensionless. This means that having
identified any relevant or marginal perturbation (as the disorder), we
find immediately another such perturbation by adding more powers of
the field. We can thus not restrict ourselves to keeping solely the
first moments of the disorder, but have to keep the whole
disorder-distribution function $R (u)$. Thus we need a {\em functional
renormalization group} treatment (FRG). Functional renormalization is
an old idea going back to the seventies, and can e.g.\ be found in
\cite{WegnerHoughton1973}, by Wegner and Houghton.  For disordered
systems, it was first proposed in 1986 by D.\ Fisher
\cite{DSFisher1986}.  Performing an infinitesimal renormalization,
i.e.\ integrating over a momentum shell \`a la Wilson, leads to the
flow $\partial _{\ell} R (u)$, with ($\epsilon =4-d$)
\begin{equation}\label{1loopRG}
\partial _{\ell} R (u) = \left(\epsilon -4 \zeta  \right) R (u) +
\zeta u R' (u) + \textstyle \frac{1}{2} R'' (u)^{2}-R'' (u)R'' (0)\ . 
\end{equation}
The first two terms come from the rescaling of $R$ and $u$
respectively. The last two terms are the result of the 1-loop
calculations, see e.g.\ \cite{LeDoussalWieseChauve2003} .

More important than the form of this equation is its actual solution,
sketched in figure \ref{fig:cusp}.
After some finite renormalization, the second derivative of the
disorder $R'' (u)$ acquires a cusp at $u=0$; the length at which this
happens is the Larkin-length. How does this overcome dimensional
reduction?  To understand this, it is interesting to study the flow of
the second and forth moment. Taking derivatives of (\ref{1loopRG})
w.r.t.\ $u$ and setting $u$ to 0, we obtain
\begin{eqnarray}
\partial_{\ell} R'' (0) &=& \left(\epsilon -2 \zeta  \right) R'' (0) +
R''' (0)^{2} \ \longrightarrow \ \left(\epsilon -2 \zeta  \right) R''
(0)\label{R2of0}\\ 
\partial_{\ell} R'''' (0) &=& \epsilon  R'''' (0) + 3 R'''' (0)^{2} +4 R'''
(0)R''''' (0)  \ \longrightarrow\ \epsilon  R'''' (0) + 3 R''''
(0)^{2}\label{R4of0} 
\ .
\end{eqnarray}\begin{figure}[t]
\centerline{\raisebox{10mm}[0mm][0mm]{\parbox{0in}{(a)}}\pfig{.4\textwidth}{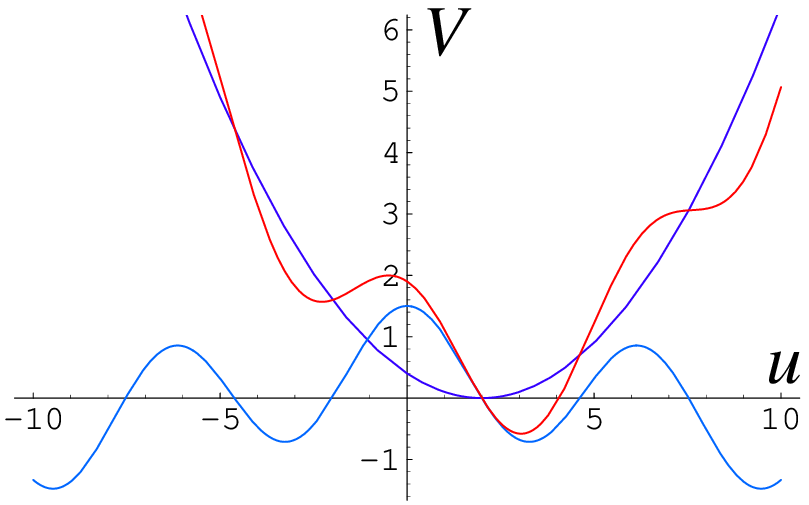}
$\quad \stackrel{\mbox{\small
minimize}}{-\!\!\!-\!\!\!-\!\!\!-\!\!\!\longrightarrow}
\quad$
\raisebox{0mm}[0mm][0mm]{\parbox{0in}{(b)}}\pfig{.4\textwidth}{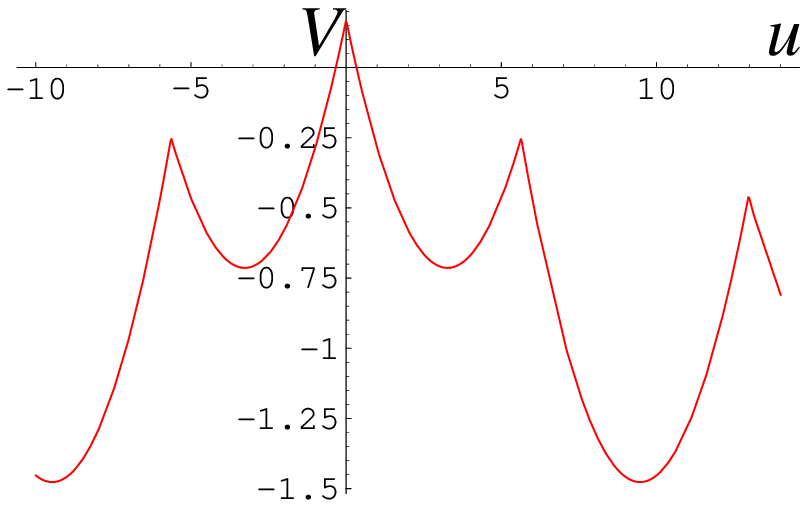}}
\centerline{\raisebox{10mm}[0mm][0mm]{\parbox{0in}{(c)}}\pfig{.4\textwidth}{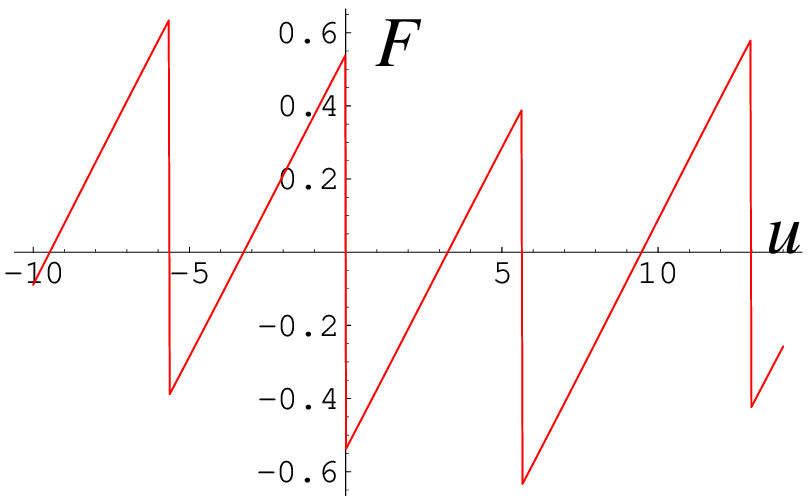}
$\quad \stackrel{\mbox{\small average}}
{-\!\!\!-\!\!\!-\!\!\!-\!\!\!\longrightarrow} \quad$
\raisebox{0mm}[0mm][0mm]{\parbox{0in}{(d)}}\pfig{.4\textwidth}{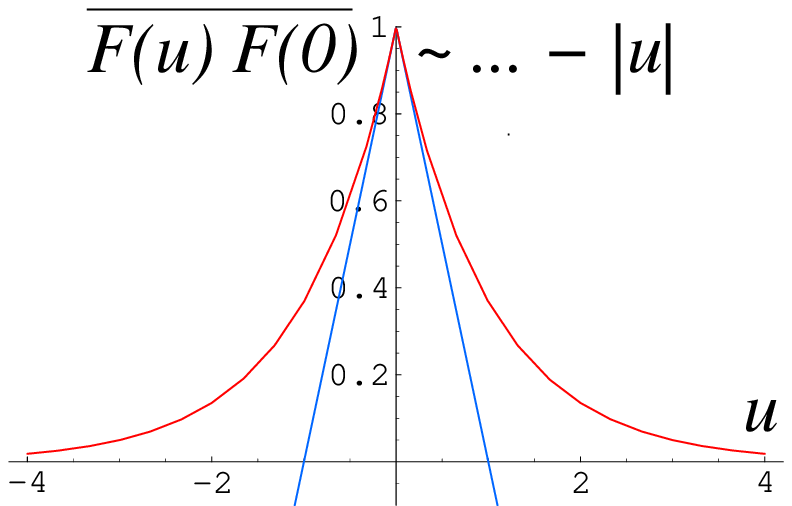}}
\caption{Generation of the cusp, as explained in the main text.} \label{minfig}
\end{figure}
Since $R (u)$ is an even function, $R''' (0)$ and $R''''' (0)$ are 0,
which we have already indicated in Eqs.\ (\ref{R2of0}) and
(\ref{R4of0}) . The above equations for $R'' (0)$ and $R'''' (0)$ are
in fact closed.  Equation (\ref{R2of0}) tells us that the flow of $R''
(0)$ is trivial and that $\zeta =\epsilon /2\equiv
\frac{4-d}{2}$. This is exactly the result predicted by dimensional
reduction. The appearance of the cusp can be inferred from equation
(\ref{R4of0}). Its solution is $R'''' (0)| _{\ell}= \frac{c\,\rme^
{\epsilon \ell }}{1-3\, c \left(\rme^ {\epsilon \ell} -1 \right)/
\epsilon },$ with $ c:= R'''' (0)| _{\ell=0}$.  Thus after a finite
renormalization $R'''' (0)$ becomes infinite: The cusp appears. By
analyzing the solution of the flow-equation (\ref{1loopRG}), one also
finds that beyond the Larkin-length $R'' (0)$ is no longer given by
(\ref{R2of0}) with $R''' (0)^{2}=0,$ but $ R''' (0)^{2} \to R'''
(0^{+})^{2} \equiv \lim_{u\to 0}R''' (u)^{2}$, which is non-zero after the
cusp.  Renormalization of the whole function thus overcomes
dimensional reduction.  The appearance of the cusp also explains why
dimensional reduction breaks down. The simplest way to see this is by
redoing the proof for elastic manifolds in disorder, which in the
absence of disorder is a simple Gaussian theory. Terms contributing to
the 2-point function involve $R'' (0)$, $TR'''' (0)$ and higher
derivatives of $R (u)$ at $u=0$, which all come with higher powers of
$T$. To obtain the limit of $T\to 0$, one sets $T=0$, and only $R''
(0)$ remains. This is the dimensional reduction result. However we
just saw that $R'''' (0)$ becomes infinite. Thus $R'''' (0) T$ may
also contribute, and the proof fails.

\section{The cusp and shocks}\label{shocks} Let us give a simple
argument of why a cusp is a physical necessity, and not an
artifact. The argument is quite old and appeared probably first in the
treatment of correlation-functions by shocks in Burgers
turbulence. It became popular in \cite{BalentsBouchaudMezard1996}. 
Suppose, we want to integrate out a single degree of freedom, whose
average position due to the elastic energy connecting it to its
neighbors is $u$. This harmonic potential and the disorder term are
represented by the parabola and the lowest curve on figure
\ref{minfig}(a) respectively; their sum is the remaining curve.  For a
given disorder realization, the minimum of the potential as a function
of $u$ is reported on figure \ref{minfig}(b). Note that it has
non-analytic points, which mark the transition from one minimum to
another.  Taking the derivative of the potential leads to the force in
figure \ref{minfig}(c). It is characterized by almost linear pieces,
and shocks (i.e.\ jumps).  Calculating the force-force correlator, the
dominant contribution in its decay for small distances is due to the
presence of shocks. Their contribution is proportional to their
probability, itself proportional to the distance between
the two observable points. This leads to $\overline{F (u)F (0)} = \overline{F
(0)^{2}} - c |u|$, with some numerical coefficient $c$.

\section{Beyond 1 loop?}\label{beyond1loop} Functional renormalization
has successfully been applied to a bunch of problems at 1-loop
order. From a field theory, we however demand more. Namely that it
allows for systematic corrections beyond 1-loop order; be
renormalizable; and thus allows to make universal predictions.
However, this has been a puzzle since 1986, and it has even been 
suggested that the theory is not renormalizable due to the appearance
of terms of order $\epsilon ^{{3}/{2}}$
\cite{BalentsDSFisher1993}. Why is the next order so complicated? The
reason is that it involves terms proportional to $R''' (0)$. A look at
figure 3 explains the puzzle. Shall we use the symmetry of $R (u)$ to
conclude that $R''' (0)$ is 0? Or shall we take the left-hand or
right-hand derivatives, related by
\begin{equation}
R''' (0^{+}) := \lim_{{u>0}\atop {u\to 0}} R ''' (u) = -
\lim_{{u<0}\atop {u\to 0}} R ''' (u) =:- R''' (0^{-}) .
\end{equation}
In the following, I will present the solution of this puzzle, at
2-loop order and large $N$.  The latter approach allows for another
independent control-parameter, and sheds further light on the
cusp-formation.

\section{Results at 2-loop order}\label{2loop} For the flow-equation
at 2-loop order, the result is
\cite{ChauveLeDoussalWiese2000a,LeDoussalWieseChauve2003,Scheidl2loopPrivate,DincerDiplom}
\begin{eqnarray}
\partial _{\ell} R (u) &=& \left(\epsilon -4 \zeta  \right) R (u) +
\zeta u R' (u) + \textstyle \frac{1}{2} R'' (u)^{2}-R'' (u)R'' (0) \nn \\
&& + \textstyle \frac{1}{2}\left(R'' (u)-R'' (0) \right)R'''
(u)^{2}-\frac{1}{2}R''' (0^{+})^{2 } R'' (u) \ . \label{2loopRG}
\end{eqnarray}
The first line is the result at 1-loop order, already given in
(\ref{1loopRG}). The second line is new. The most interesting term is
the last one, which involves $R''' (0^{+})^{2}$ and which we therefore
call {\em anomalous}.  The hard task is to fix the prefactor
$(-\frac{1}{2})$.  We have found five different prescriptions to
calculate it: The sloop-algorithm, recursive construction,
reparametrization invariance, renormalizability, and potentiality
\cite{ChauveLeDoussalWiese2000a,LeDoussalWieseChauvePREPa}. For lack
of space, we restrain our discussion to the last two ones. At 2-loop
order the following diagram appears
\begin{equation}\label{rebi}
\pfig{2cm}{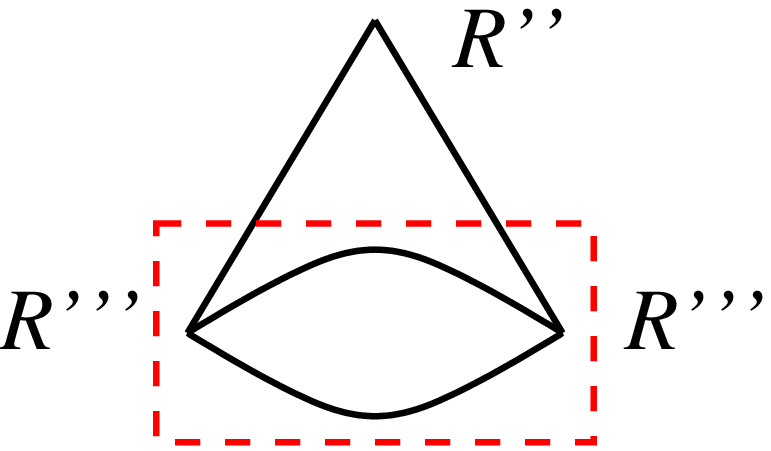}\  \longrightarrow\ \frac{1}{2}\left(R'' (u)-R'' (0)
\right)R''' 
(u)^{2} -\half R'' (u)R''' (0^{+})^{2}
\end{equation}
leading to the anomalous term. The integral (not written here)
contains a subdivergence, which is indicated by the
box. Renormalizability demands that its leading divergence (which is
of order $1/\epsilon ^{2}$) be canceled by a 1-loop counter-term. The
latter is unique thus fixing the prefactor of the anomalous term. 

Another very physical demand is that the problem remain potential,
i.e.\ that forces still derive from a potential. The force-force
correlation function being $-R'' (u)$, this means that the flow of
$R' (0)$ has to be strictly 0.  From (\ref{2loop}) one can check that
this does not remain true if one changes the prefactor of the last
term in (\ref{2loop}); thus fixing it.

Let us give some results for random-bond disorder (short-ranged
potential-potential correlation function). For this, we have to solve
(\ref{2loopRG}) numerically, with the result $\zeta = 0.208 298 04
\epsilon +0.006858 \epsilon ^{2}$. This compares well with numerical
simulations, see figure \ref{fig:numstat}.

\begin{figure}[tb]\centerline{\small
\begin{tabular}{|c|c|c|c|c|}
\hline
$\zeta _{\rm}$ & one loop & two loop & estimate & 
simulation and exact\\
\hline
\hline
$d=3$  & 0.208 &  0.215  & $0.215\pm 0.01$  & 
$0.22\pm 0.01$ \cite{Middleton1995}  \\
\hline
$d=2$ &0.417 &0.444 &$0.42\pm 0.02$ &  $0.41\pm 0.01$ \cite{Middleton1995} \\
\hline
$d=1$ & 0.625 & 0.687 &  $0.67\pm 0.02$ & $2/3$ \\
\hline
\end{tabular}}\medskip 
\caption{Results for $\zeta $ in the random bond case.}\label{fig:numstat}
\end{figure}

\section{Finite $N$}\label{s:finiteN} Up to now, we have studied the
functional RG in two cases: For one component $N=1$ and in the limit
of a large number of components, $N\to \infty$. The general case of
finite $N$ is more difficult to handle, since derivatives of the
renormalized disorder now depend on the direction, in which this
derivative are taken. Define amplitude $u:=|\vec u|$ and direction
$\hat u:= \vec u/|\vec u|$ of the field. Then deriving the latter
variable leads to terms proportional to $1/u$, which are diverging in
the limit of $u\to 0$. This poses additional problems in the
calculation, beyond the case $N=1$. At 1-loop order everything is
well-defined \cite{BalentsDSFisher1993}. We have found a consistent
RG-equation at 2-loop order (see \cite{Wiese2003a} and unpublished):
\begin{figure}[bt] 
\centerline{\scalebox{.8}{{\unitlength1mm
\begin{picture} (90,55)
\put(0,0){\fig{85mm}{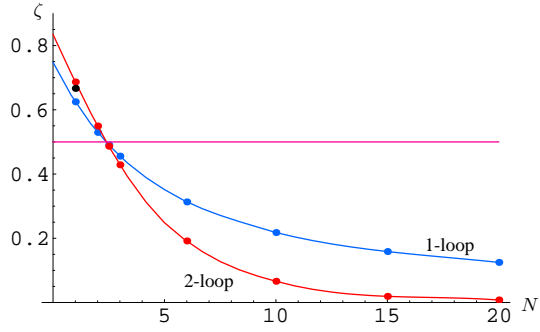}}
\put(5,52){$\zeta $}
\put(86,3){$N$} 
\put(70,13){1-loop}
\put(30,7){2-loop}
\end{picture}}}
} 
\caption{Results for the roughness $\zeta$ at 1- and 2-loop order, as
a function of the number of components $N$.}  \label{f:Ncomp}
\end{figure}

The fixed point equation has to be integrated numerically, order by
order in $\epsilon$. The result, specialized to directed polymers,
i.e.\ $\epsilon =3$ is plotted on figure \ref{f:Ncomp}.  We see that
the 2-loop corrections are rather big at large $N$, so some doubt on
the applicability of the latter down to $\epsilon=3$ is
advised. However both 1- and 2-loop results reproduce well the two
known points on the curve: $\zeta =2/3$ for $N=1$ and $\zeta =0$ for
$N=\infty$. The latter result has been given in section \ref{largeN}
Via the equivalence \cite{KPZ} of the directed polymer problem in $N$
dimensions treated here and the KPZ-equation of non-linear surface
growth in $N$ dimensions, we conclude that $d\approx 2.4$ is the
upper critical dimension of KPZ.

\section{Large $N$}\label{largeN} In the last section, we have
discussed renormalization in a loop expansion, i.e.\ expansion in
$\epsilon$. In order to independently check consistency it is good to
have a non-perturbative approach. This is achieved by the large-$N$
limit, which can be solved analytically and to which we turn now. We
start from the disorder-averaged energy with disorder correlator $B
(\vec u^{2}) \equiv R (|\vec u|)$ where we
use an $N$-component field $\vec{u} $.  We then calculate the
 free energy in presence of a source $j$, and finally the effective
action $\Gamma (\vec u) $ via a Legendre transform. For large $N$ the
saddle point equation reads \cite{LeDoussalWiese2001}
\begin{equation}\label{saddlepointequation}
\tilde B' (u_{ab}^{2}) = B' \left( \chi_{ab}\right)\ , \qquad \chi_{ab} =
u_{ab}^{2}+2 T I_{1} + 4 I_{2} [\tilde B' (u_{ab}^{2})-\tilde B' (0)]
\end{equation}
This equation gives the derivative of the effective (renormalized)
disorder $\tilde B$ as a function of the (constant) background field
$u_{ab}^{2}= (u_{a}-u_{b})^{2}$ in terms of: the derivative of the
microscopic (bare) disorder $B$, the temperature $T$ and the integrals
$I_{n}:= \int_{k}\frac{1}{\left(k^{2}+m^{2} \right)^{n}}$.

The saddle-point equation can be turned into a closed functional
renormalization group equation for $\tilde B$ by taking the derivative
w.r.t.\ $m$ (restricting ourselves to $T=0$):
\begin{equation}    \hspace{-1.5 cm}
\partial _{l}\tilde B (x)\equiv -\frac{m \partial }{\partial m}\tilde
B (x) =\left(\epsilon -4\zeta \right)\! \tilde B (x) + 2 \zeta x
\tilde B' (x)+\frac{1}{2}\tilde B' (x)^{2}-\tilde B' (x) \tilde B'
(0)
\end{equation}
This is a complicated nonlinear partial differential equation. It is
therefore surprising, that one can find an analytic solution. (The
trick is to write down the flow-equation for the inverse function of
$\tilde B' (x)$, which is linear.) Let us only give the results of
this analytic solution: First of all, for long-range correlated
disorder of the form $\tilde B' (x)\sim x^{-\gamma }$, the exponent
$\zeta $ can be calculated analytically as $\zeta =\frac{\epsilon }{2
(1+\gamma )}\ . $ It agrees with the replica-treatment in
\cite{MezardParisi1991} and the 1-loop treatment in
\cite{BalentsDSFisher1993}. For short-range correlated disorder,
$\zeta =0$.  Second, it demonstrates that before the Larkin-length,
$\tilde B (x)$ is analytic and thus dimensional reduction
holds. Beyond the Larkin length, $\tilde B'' (0)=\infty $, a cusp
appears and dimensional reduction is incorrect. This shows again that
the cusp is not an artifact of the perturbative expansion, but an
important property even of the exact solution of the problem (here in
the limit of large $N$).

\section{Relation to Replica Symmetry Breaking (RSB)}\label{s:RSB} There
is another treatment of the limit of large $N$ given by M\'ezard and
Parisi \cite{MezardParisi1991}. They make  a
Gaussian variational ansatz of the form
\begin{eqnarray}\label{HlargeNMP}
{\cal H}_{\mathrm g}[\vec u] &=& \frac{1}{2T} \sum _{a=1}^{n}\int_{x} 
 \vec u_{a} (x)\left(-\nabla^{2}{+}m^{2} \right) \vec u_{a} (x) 
   -\frac{1}{2 T^{2}}  \sum
_{a,b=1}^{n} \sigma_{ab} \, \vec u_{a} (x)\vec u_{b} (x)\ ,
\end{eqnarray}
which becomes exact for $N\to\infty$. The art is to make an
appropriate ansatz for $\sigma_{ab}$. The simplest possibility,
$\sigma _{ab}=\sigma $ for all $a\neq b$ reproduces the dimensional
reduction result, which breaks down at the Larkin length. Beyond that
scale, a replica symmetry broken (RSB) ansatz for $\sigma _{ab}$ is
necessary, of the form
$\sigma_{ab} =
\left(\,\pfig{1cm}{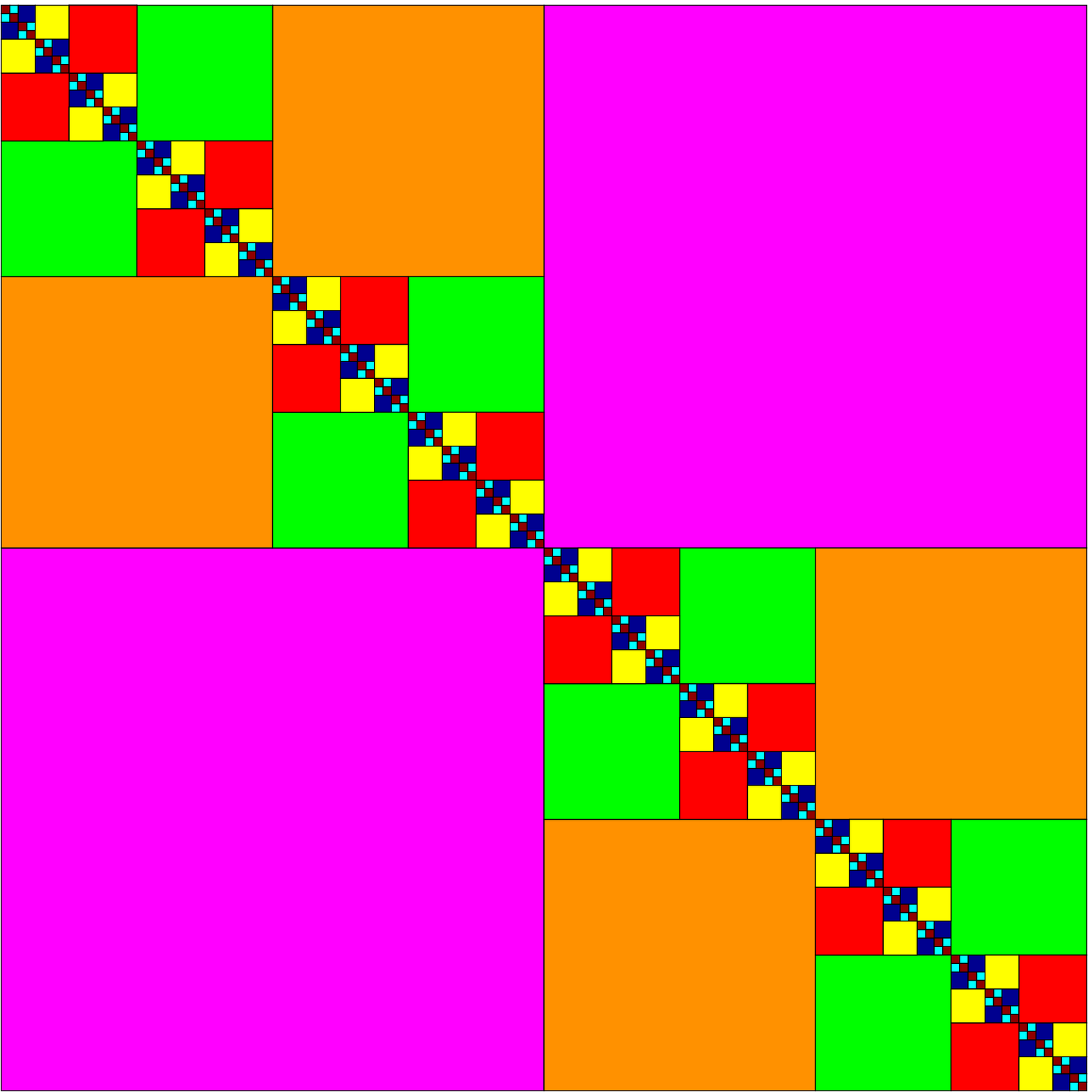}\,\right)\ .$
\begin{figure}[b]\label{6}
\centerline{\fig{6cm}{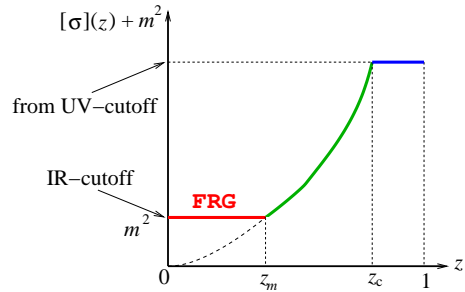}}
\caption{The function $\left[\sigma \right] (u)+m^{2}$ as given in
\protect\cite{MezardParisi1991}.} \vspace{-0.1cm}\label{fig:MP-function}
\end{figure}Parisi has shown that this infinitely often
replica-symmetry broken matrix can be parameterized by a function
$[\sigma] (z)$ with $z\in \left[0,1 \right]$ where $z=0$ describes
distant states, whereas $z=1$ describes nearby states. The solution of
the large-$N$ saddle-point equations leads to the curve depicted in
figure \ref{6} Knowing it, the 2-point function is given by $\left<
u_{k}u_{-k} \right>=\frac{1}{k^{2}+m^{2}}\left(1+\int_{0}^{1}
\frac{\rmd z}{z^{2}} \frac{\left[\sigma \right]
(z)}{k^{2}+\left[\sigma \right] (z)+m^{2}} \right).$
 
What is the relation between the two
approaches, which both pretend to calculate the same 2-point function?
Comparing the analytical solutions, we find that the 2-point function
given by FRG is the same as that of RSB, if in the latter expression
we only take into account the contribution from the most distant
states, i.e.\ those for $z$ between 0 and $z_{m}$ (see figure
\ref{fig:MP-function}). To understand why this is so, we have to
remember that the two calculations were done under quite different
assumptions: In contrast to the RSB-calculation, the FRG-approach
calculated the partition function in presence of an external field
$j$, which was then used to give via a Legendre transformation the
effective action. Even if the field $j$ is finally turned to 0, the
system will remember its preparation, as is the case for a magnet.

 By explicitly breaking the replica-symmetry through an
applied field, all replicas will settle in distant states, and the
close states from the Parisi-function $\left[\sigma \right] (z)+m^{2}$
(which describes {\em spontaneous} RSB) will not contribute.  However,
we found that the full RSB-result can be reconstructed by remarking
that the part of the curve between $z_{m}$ and $z_{c}$ is independent
of the infrared cutoff $m$, and then integrating over $m$
\cite{LeDoussalWiese2001} ($m_{c}$ is the mass corresponding to
$z_{c}$):
\begin{equation}\label{RSB=intFRG}
\left< u_{k}u_{-k} \right>\Big|^{\mathrm{RSB}}_{k=0} =\frac{\tilde
R'_{m}(0)}{m^{4}} +\int_{m}^{m_{c}} \frac{\rmd \tilde
R'_{\mu}(0)}{\mu^{4}} + \frac{1}{m_{c}^{2}}-\frac{1}{m^{2}}\ .
\end{equation}
We also note that a similar effective action  has been proposed in
\cite{BalentsBouchaudMezard1996}. While it agrees qualitatively, it
does not reproduce the correct FRG 2-point function, as
it should.

\section{Corrections at order $1/N$}\label{sec:1overN}
In a graphical notation, we find \cite{LeDoussalWiese2004a}
\begin{eqnarray}
\delta B^{(1)}&=&
\!\!\diagram{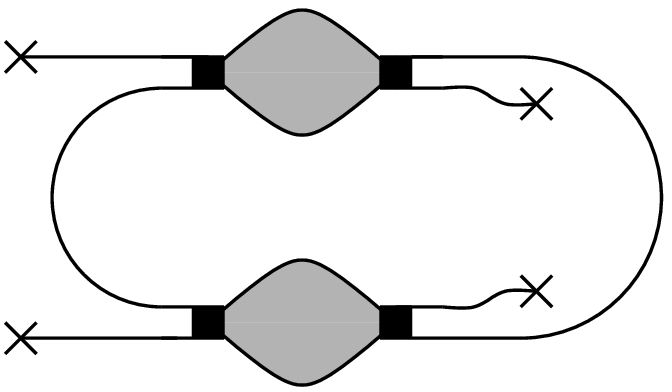}\!\!+\!\!\!\diagram{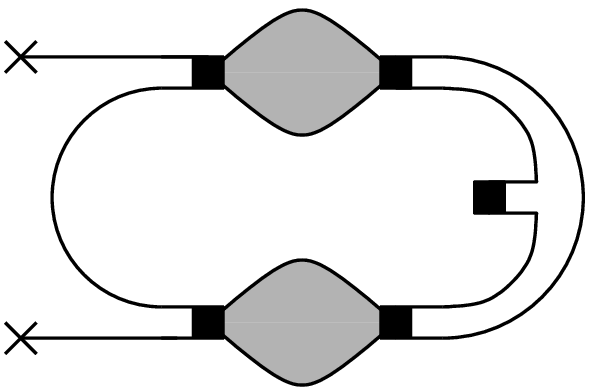}\!\!+\!\!\diagram{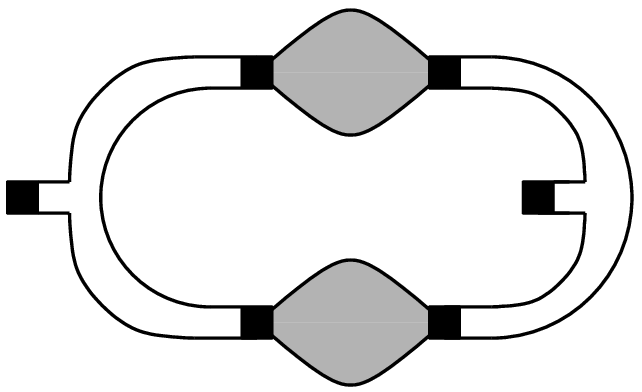}\!\!+\!\!\!\diagram{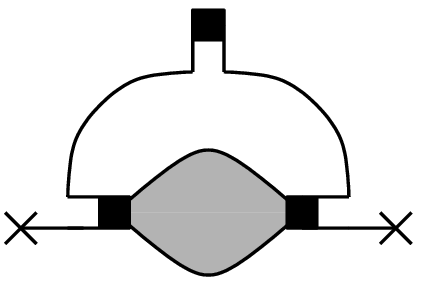}\!\!+\!\!\diagram{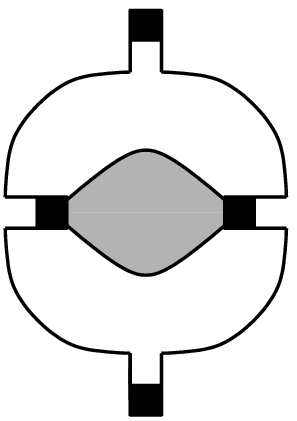}\!\! \nonumber 
\\
&& +T\Big( \!\!\diagram{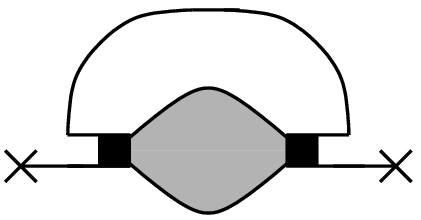} \!\!+ \!\!\diagram{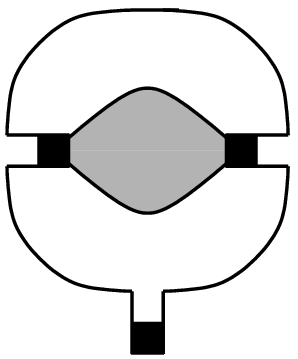} \!\!+
\!\!\diagram{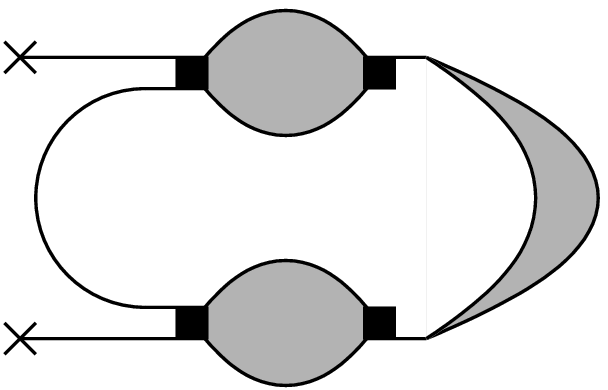} \!\!+ \!\!\diagram{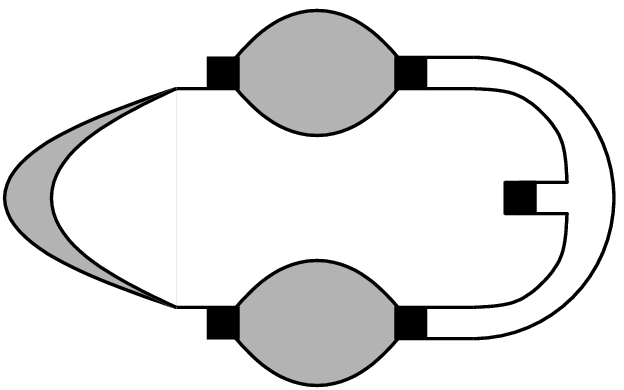}\!\!+
\!\!\diagram{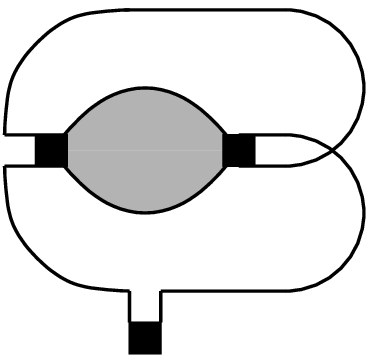} \!\!+ \!\!\diagram{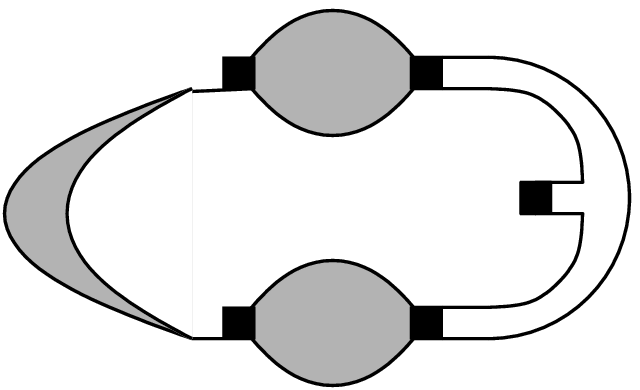}\!\! \Big)\nonumber \\
&& + T^{2}\Big( \!\!\diagram{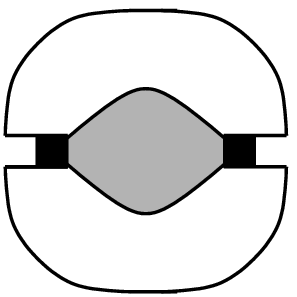} \!\!+ \!\!\diagram{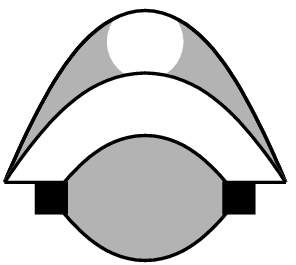}
\!\!+
\!\!\diagram{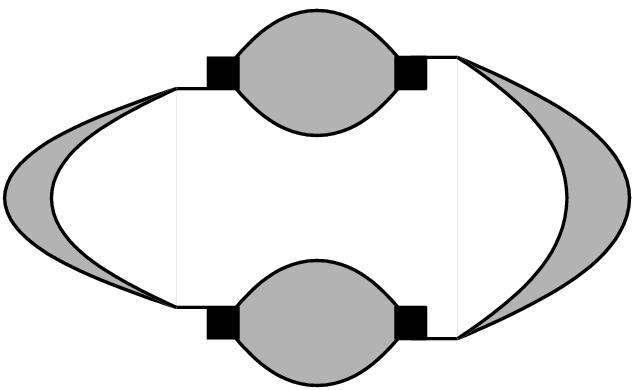} + {\cal A}^{T^{2}}\Big)\\
\diagram{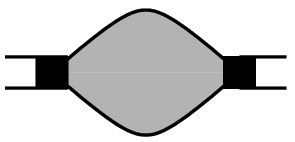}&=&B'' (\chi _{ab})\left(1-4A_{d} I_{2} (p)B'' (\chi
_{ab}) \right)^{-1}\ ,\quad  \diagram{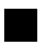}=B(\chi_{ab})\ ,
\end{eqnarray}
where the explicit expressions are given in
\cite{LeDoussalWiese2004a}. 
By varying the IR-regulator, one can derive a $\beta$-function at
order $1/N$, see \cite{LeDoussalWiese2004a}. At $T=0$, it is
UV-convegent, and should allow to find a fixed point. We have been
able to do this at order $\epsilon$, showing consistency with the
1-loop result, see section \ref{s:finiteN}. Other dimensions are more
complicated. 

A $\beta$-function can also be defined at finite $T$. However since
temperature is an irrelevant variable, it makes the theory
non-renormalizable, i.e.\ in otder to define it, one must keep an
explicit infrared cutoff. These problems will be treated in a
forthcoming publication. 

\section{Perspectives}\label{perspectives} Other interesting problems
have been treated by the above methods, especially dynamic problems
(see \cite{Wiese2003a} for a review); and many more are
now in reach. Some open points have already been raised in these
notes, others are the strong disorder phase of random field problems,
or wether FRG can also be applied to spin-glasses. We have to leave
these problems for future research and as a challenge for the reader
to plunge deeper into the mysteries of functional renormalization.

\subsection*{Acknowledgements}\label{ack} It is a pleasure to thank
the organizers of StatPhys22 for the opportunity to give this
lecture. The results presented here have been obtained in a series of
inspiring ongoing collaborations with Pierre Le Doussal, and I am
grateful to him and my other collaborators Pascal Chauve, Werner
Krauth and Alberto Rosso for all their enthousiasm and dedicated
work. Numerous discussions with Leon Balents, Edouard Br\'ezin,
Andreas Ludwig Thomas Nattermann and Stefan Scheidl are gratefully
acknowledged.


\end{document}